\documentstyle[aps,preprint,prb]{revtex}
\begin{document}
\title{\bf Impurity scattering in superconductors$^{\dagger}$}
\author{Yong-Jihn Kim and K. J. Chang}
\address{Department of Physics,  Korea Advanced Institute of Science and 
Technology,\\
 Taejon 305-701, Korea}
\maketitle
\begin{abstract}
We present some of the recent theoretical studies
on the impurity effects in conventional superconductors, such as 
magnetic  and ordinary impurity effects in dirty and weak localization
limits, and describe successfully unexplained experimental results.
We find that the critical sheet resistance for the suppression of
superconductivity in thin films depends on superconductor,
and point out that impurity dopings in high $T_{c}$ superconductors
cause a metal-insulator transition and thereby suppress $T_{c}$.

\end{abstract}
\vskip 4pc
\noindent
$^{\dagger}$ Invited Talk at the Symposium 97 on Theoretical Solid State Physics

February 3-5, 1997,  Taejon, Korea
\vskip 0.1pc
\noindent
$^{\dagger}$ 
To appear in the Journal of Korean Physical Society Supplement
\vskip 3.0pc
\noindent

\vfill\eject
\section{Introduction}
Although there have been much studies on the impurity doping effects in high $T_{c}$
superconductors,$^{1,2}$ it has not been successful to understand the underlying
physics. In particular, the decrease of $T_{c}$ due to impurity scattering
is not fast enough to support the anisotropic pairing model, such as d-wave 
or anisotropic s-wave pairing. In conventional low $T_{c}$ superconductors,
it was shown that the Abrikosov and Gor'kov's (AG) Green function
theory is in conflict with the Anderson's theory of dirty 
superconductors.$^{3-5}$ The Anderson's theory predicted no substantial
decrease of $T_{c}$ with ordinary impurity substitution, while the AG theory 
leads to a large decrease of $T_{c}$, which varies linearly with 
impurity concentration. For magnetic impurity effects, the $T_{c}$ 
reduction caused by exchange scattering was found to be suppressed
by adding non-magnetic impurities or radiation damage,$^{6,7}$
which contradicts to the AG theory. In weak localization limit,
ordinary impurities also decrease $T_{c}$, which is beyond the Anderson's theory.
Previously, this feature was attributed to the enhanced Coulomb repulsion
due to impurities,$^{8,9}$ however, tunneling experiments did not
show any increase of the Coulomb repulsion.$^{10,11}$  

In this paper, we present some of our recent theoretical work$^{3, 12-15}$
on the impurity effects in conventional superconductors, which describe
successfully unexplained experimental results. We first discuss
the magnetic impurity effect and show why ordinary impurities and 
radiation damage suppress the $T_{c}$ reduction caused by magnetic impurities
within the framework of the BCS theory. The ordinary impurity effect is 
considered both in dirty and weak localization limits. In weak localization
limit, where the Anderson's theorem is not valid, $T_{c}$ is suppressed
because of the amplitude modulation of the wavefunction caused by
impurity scattering. In two-dimensional samples, it is shown that
the critical sheet resistance for the suppression of superconductivity
depends on superconducting material, in good agreement
with experiments.$^{16,17}$ We point out that the impurity doping and
ion-beam-induced damage in high $T_{c}$ superconductors give rise to a 
metal-insulator transition and thereby suppress $T_{c}$.

\section{Impurity scattering in conventional superconductors}
\subsection{Magnetic impurity scattering}

Magnetic impurities suppress strongly the transition temperatures of
singlet-pairing superconductors. The magnetic interaction between
a conduction electron at $\bf r$ and a magnetic solute at ${\bf R}_{i}$
is given by 
\begin{eqnarray}
H_{mag}=\sum_{i}J{\vec s}\cdot{\vec S}_{i}\delta({\bf r}-{\bf R}_{i}),
\end{eqnarray}
where ${\vec s}={1\over 2}{\vec \sigma}$ and $\vec \sigma$ represents
the Pauli spin operators. Employing the degenerate scattered-state pairs 
with the magnetic scattering effect, Kim and Overhauser$^{12}$ showed 
that the phonon-mediated matrix elements are expressed as
\begin{eqnarray}
V_{nn'}=-V\langle cos\theta_{n'}({\bf r})cos\theta_{n}({\bf r})\rangle_{av},
\end{eqnarray}
where $\langle \ \rangle_{av}$ denotes the average over ${\bf r}, {\vec R}_{i}$
and the spin direction ${\vec S}_{i}$ of the solutes.
Here $cos\theta({\bf r})$ is the relative singlet amplitude of the basis
pair when both the electrons are near $\bf r$. In the case of magnetic 
scattering, the singlet amplitude of the Cooper pair is reduced, and 
consequently $T_{c}$ is decreased.

Notice that only the magnetic impurities within the BCS coherence distance
$\xi_{o}$ (for a pure superconductor) from the Cooper pair's center of
mass reduce both the singlet amplitude and the pairing interaction.
Then, one finds
\begin{eqnarray}
\langle cos\theta\rangle = 1- {\pi\xi_{o}\over 2\ell_{s}},
\end{eqnarray}
where
\begin{eqnarray}
\xi_{o}=0.18{\hbar v_{F}\over k_{B}T_{c}},
\end{eqnarray}
and $\ell_{s}$ is the mean free path for the exchange scattering only and
$v_{F}$ is the Fermi velocity.
Here, we use an iterative method to calculate $T_{c}$ from the 
BCS gap equation.
The change of $T_{c}$ with respect to $T_{co}$ (for a pure metal)
can be easily calculated to the first order
in impurity concentration, such as
\begin{eqnarray}
k_{B}\Delta T_{c} \cong -{0.18\pi\hbar \over \lambda\tau_{s}},
\end{eqnarray}
where $\lambda$ is the electron-phonon coupling constant and $\tau_{s}$ is the spin-disorder scattering time.
Since $\Delta T_{c}$ is inversely proportional to $\lambda$, the initial
slope (versus $1/\tau_{s}$) depends on superconductor, indicating that
this slope is not the universal constant, while the universal
behavior was proposed by the AG theory. As illustrated in Fig. 1,
the critical temperatures of weakly-coupled superconductors are found to 
decrease rapidly with increasing of magnetic impurities, as compared to
strongly-coupled superconductors. 

For the conduction electrons with a mean free path $\ell$, which is smaller
than $\xi_{o}$, the effective coherence length is reduced to
\begin{eqnarray}
\xi_{eff}\approx \sqrt{\ell\xi_{o}}.
\end{eqnarray}
Then, the reduction of $T_{c}$ by magnetic impurities is significantly
suppressed because the number of magnetic impurities within the distance of
$\xi_{eff}$ is very small.
This behavior was first observed by co-doping with non-magnetic
impurities that result in the decrease of $\ell$.$^{6}$
A similar compensation effect was also observed by radiation damage;
for pure and Mn-implanted In-films, the change of $T_{c}$ were found to be
2.2 and 0.3 K, respectively, after an exposure to a 275 kev 
$\rm Ar^{+}$-ion fluence of 2.2$\times 10^{16}cm^{-2}$.$^{7}$
This compensation effect contradicts to the traditional belief
that $\Delta T_{c}$ in magnetically doped superconductors
is unaffected by non-magnetic scatterings.$^{4}$

\subsection{Ordinary impurity scatterings in dirty and weak localization
limits}

To describe the ordinary impurity effect, Anderson introduced
the exact scattered states $\psi_{n\sigma}$ for the conduction
electrons in a metal with ordinary impurities,
which form time-reversed scattered state pairs.
The scattered state $\psi_{n\sigma}$ can be expanded
in terms of plane waves $\phi_{{\vec k}\sigma}$, such as
\begin{eqnarray}
\psi_{n\sigma}=\sum_{\vec k}\phi_{{\vec k}\sigma}\langle {\vec k}|
n\rangle.
\end{eqnarray}
Then, the phonon-mediated matrix elements between the time-reversed
pairs are written as$^{3}$
\begin{eqnarray}
V_{nn'}=-V(1+\sum_{{\vec k}\not= -{\vec k'},{\vec q}}\langle -{\vec k'}|
n\rangle \langle{\vec k}|n\rangle^{*} \langle {\vec k}-{\vec q}|n'\rangle
\langle -{\vec k'}-{\vec q}|n'\rangle^{*}).
\end{eqnarray}
Although the correction term in the right hand side of Eq. (8) is 
negligibly small in dirty limit, it is important in weak localization limit.
Consequently, the Anderson's theorem is valid only in dirty limit, to
the first order in impurity concentration.$^{3}$

It was shown$^{13}$ that the Tsuneto's strong coupling theory$^{18}$
fails to explain the existence of the localization correction
in the phonon-mediated interaction. Alternatively, from the
real space formalism of the strong coupling theory with the time-reversed
pairs, Kim$^{13}$ obtained a strong coupling gap equation
\begin{eqnarray}
\Delta(n,\omega)=\sum_{\omega'}\lambda(\omega-\omega')\sum_{n'}V_{nn'}
{\Delta(n',\omega')\over \omega'^{2}+\epsilon_{n'}^{2}},
\end{eqnarray}
where
\begin{eqnarray}
V_{nn'}=-V\int|\psi_{n}({\bf r})|^{2}|\psi_{n'}({\bf r})|^{2}d{\bf r},
\end{eqnarray}
\begin{eqnarray}
\lambda(\omega-\omega')={\omega_{D}^{2}\over \omega_{D}^{2}+(\omega-\omega')^{2}},
\end{eqnarray}
and $\omega_{D}$ is the Debye frequency. Here $\omega'$ and $\epsilon_{n'}$
represent the Matsubara frequency and the electron energy, respectively.
Using the wavefunction in Eq. (7), one can easily derive the formula
of Eq. (8) from Eq. (10). We point out that $V_{nn'}$
gives the change of the phonon-mediated interaction due to impurities, and it 
decays exponentially for the localized states. In the Tsuneto's theory,
however, it remains unchanged, even if the wavefunctions are localized.

For the strongly localized states, since both the phonon-mediated 
interaction and the conductivity decay exponentially, they are expected to
have the same correction term in weak localization limit. Kaveh 
and Mott$^{19}$ showed that the wavefunction for the weakly localized states
may be written as a mixture of power-law and extended wavefuctions.
Employing their wavefunctions and considering only the impurities 
within the distance of $\xi_{eff}$, the following relations for 
the matrix elements were obtained;$^{14,15}$
\begin{eqnarray}
V_{nn'}^{3d} &\cong& 
-V[1-{3\over (k_{F}\ell)^{2}}(1-{\ell\over L})], \\
V_{nn'}^{2d} &\cong&  
 -V[1-{2\over \pi k_{F}\ell}ln(L/\ell)], \\
V_{nn'}^{1d} &\cong&  
 -V[1-{1\over (\pi k_{F}a)^{2}}(L/\ell-1)],
\end{eqnarray}
where $\ell$ and $L$ are the elastic and inelastic mean free
paths, respectively, and $a$ is the radius of an one-dimensional
wire.

Solving the BCS gap equation with the matrix element of
Eq. (12), the change of $T_{c}$ with respect to $T_{co}$ 
satisfies the relation
\begin{eqnarray}
{T_{co}-T_{c}\over T_{co}} \propto {1\over (k_{F}\ell)^{2}},
\end{eqnarray}
for bulk materials and this result is in good agreement with 
experiments,$^{20}$ as shown in Fig. 2. In homogeneous 
two-dimensional thin films, an empirical formula was 
obtained,$^{21}$
\begin{eqnarray}
{T_{co}-T_{c}\over T_{co}}\propto {1\over k_{F}\ell}\propto R_{sq},
\end{eqnarray}
where $R_{sq}$ denotes the sheet resistance, and in fact this 
formula can be derived by putting Eq. (13) into the BCS gap 
equation. Previously, the decrease of $T_{c}$ due to disorder
was attributed to the enhanced Coulomb repulsion,$^{8,9}$
whereas tunneling measurements do not support this 
picture.$^{10,11}$ The dirty boson theory$^{22}$ predicted the 
universal critical sheet resistance of $R_{q}=h/4e^{2}=6.4k\Omega$
for the suppression of superconductivity in thin films. However,
we find that from Eq. (13) the critical sheet resistance is not
universal but sample-dependent, which agrees with experiments.$^{17,23}$

\section{Impurity scattering in high $T_{c}$ superconductors}

There is currently considerable interest in the symmetry of the 
superconducting state in high $T_{c}$ superconductors.$^{24}$
Impurity doping studies may give a clue of resolving this 
problem. Since the normal-state transport behavior
of the cuprates is so anomalous, the impurity doping effects on 
the superconducting state may not be easy to understand.
We note that almost all the experiments with Cu ions substituted by
other metal ions show the metal-insulator transition driven
by impurity dopings.$^{25,26}$ Thus, the decrease of $T_{c}$ 
seems to be closely related to the wavefunction localization.
If the d-wave pairing and the Fermi liquid theory are assumed in 
the cuprates, since $T_{c}$ decreases much faster with increasing 
of impurities,$^{12}$ the superconductivity may disappear before
the metal-insulator transition is reached. In ion-beam irradiation
and ion implantation experiments$^{27,28}$ as well as  
$\rm Y_{1-x}Pr_{x}Ba_{2}Cu_{3}O_{7}$ samples,$^{29,30}$ the 
metal-insulator transitions were also found in the doping region, 
where $T_{c}$ drops to zero.
This anomalous impurity doping effect seems to imply that the 
Landau's quasi-particle picture is not applicable for the cuprates. 
To understand this anomalous behavior, we may need to understand
the impurity effect on the normal state.

Recently, Suryanarayanan et al.$^{31}$ found the recovery of 
superconductivity in $\rm Y_{1-x}Ca_{x}SrBaCu_{2.6}Al_{0.4}O_{6+z}$  
when Y is substituted by Ca; the values of $T_{c}$ were found to be
0, 29, and 47 K for $\rm x=0,\ 0.1\,$ and 0.2, respectively.
This results may be understood if we consider the importance of the 
metal-insulator transition caused by impurity doping. 
When ${\rm x}=0$, the increase of Al impurities 
changes the system to the insulating state with the mobility edge
lying below the Fermi energy. Then, since the conducting electrons 
are localized, the superconductivity transition does not appear.
If holes are added in the Cu-O planes with Ca impurities, 
since the Fermi energy moves below the mobility edge, assuming
the mobility edge unchanged, and the electrons are extended,
Ca-doped systems become superconducting.
Further experimental studies are needed to understand clearly the 
impurity-driven metal-insulator transition in the cuprates.
 
\section{Conclusions}
Within the framework of the BCS theory, we have discussed the magnetic 
and non-magnetic impurity effects on conventional superconductors.
In particular, we find that the critical sheet resistance for the
suppression of superconductivity in thin films is not universal
but sample-dependent. For high $T_{c}$ superconductors, it is pointed
out that the metal-insulator transition driven by impurity doping is 
important in understanding both the normal and superconducting states 
of the cuprates.

\centerline{\bf ACKNOWLEDGMENTS}
YJK is grateful to A. W. Overhauser and Yunkyu Bang for discussions.
This work has been supported by the Brain pool project of KOSEF and
the MOST.

\vfill\eject

\centerline{\bf Figure Captions}
\vskip 1pc
\noindent
Fig.1 Initial slopes of $T_{c}$ with varying the spin-disorder scattering
rate (1/$\tau_{s}$) for $T_{c}=$1, 5, 

and 15K.
\vskip 1pc
\noindent
Fig.2 The calculated superconducting temperatures (solid line) of
$\rm InO_{x}$ are plotted 

as a function of $(k_{F}\ell)^{-2}$ and compared with
experiments (triangles) from Ref. 20.
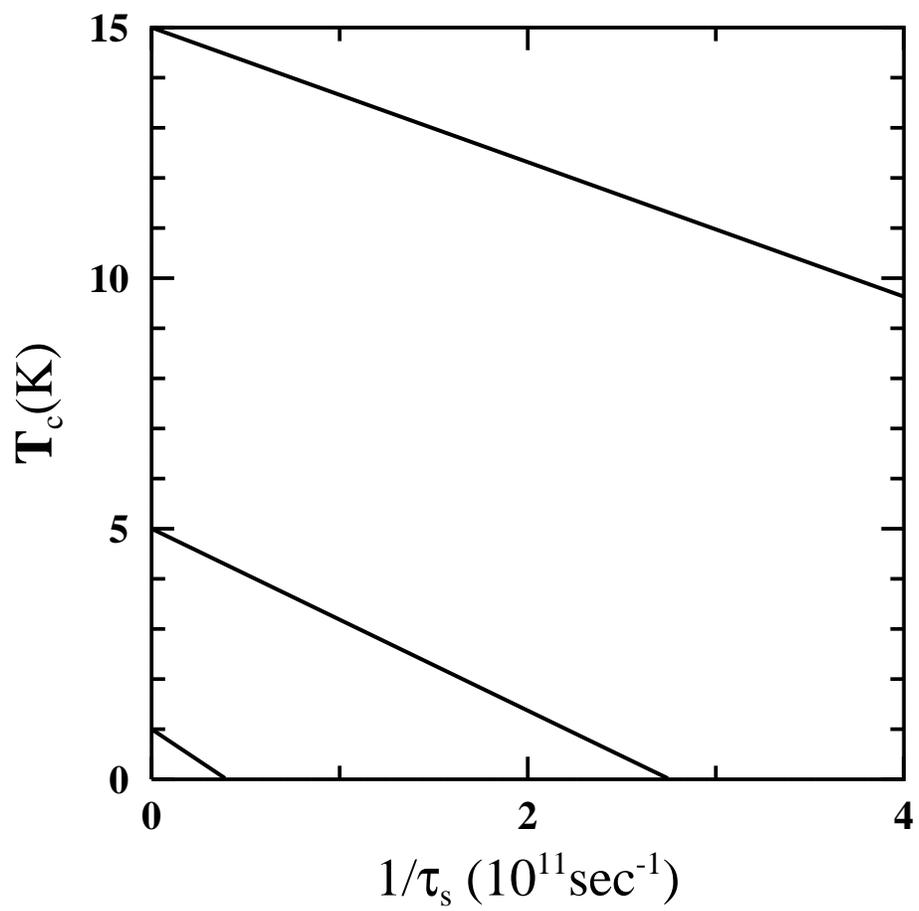
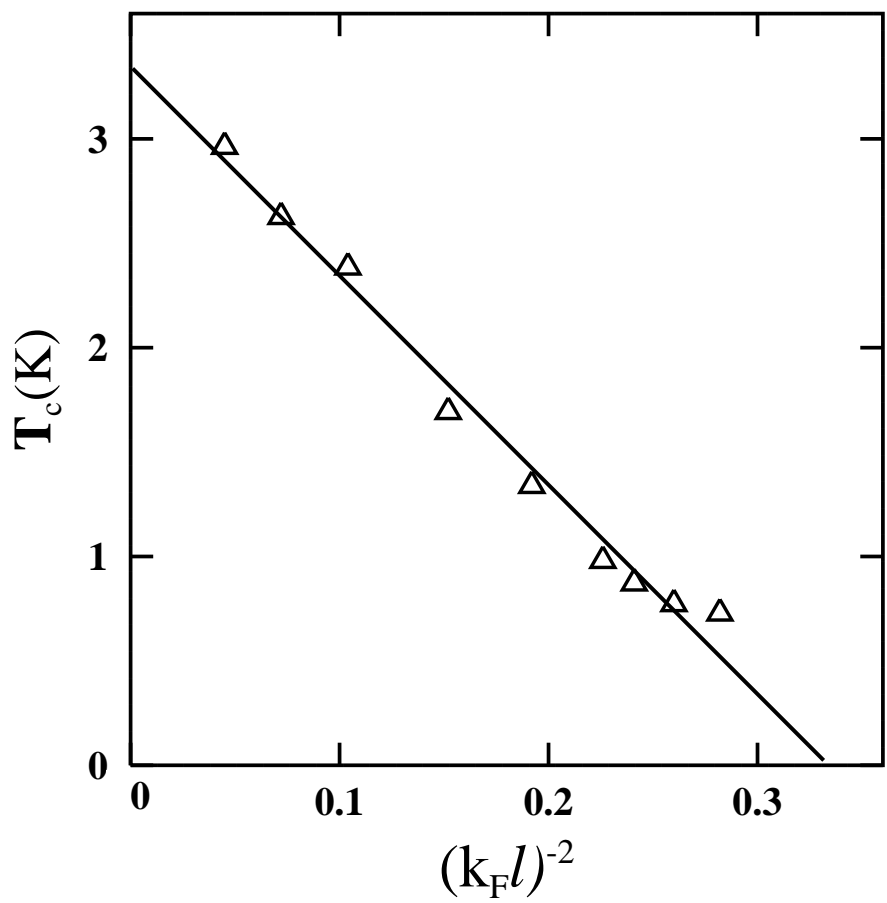
\end{document}